\documentclass[aps,preprint,showpacs,amsmath,amssymb,superscriptaddress]{revtex4-2}
\usepackage{graphicx}
\usepackage{dcolumn}
\usepackage{bm}
\usepackage[colorlinks=true, linkcolor=blue, citecolor=blue, urlcolor=blue]{hyperref}

\begin{document}
%\preprint{APS/123-QED}

\title{Dominant apical-oxygen electron-phonon coupling in HgBa$_2$Ca$_2$Cu$_3$O$_{8+\delta}$}

\author{Wenshan~Hong}
\thanks{These authors contributed equally to this work}
\affiliation{Beijing National Laboratory for Condensed Matter Physics, Institute of Physics, Chinese Academy of Sciences, Beijing, China}
\affiliation{International Center for Quantum Materials, School of Physics, Peking University, Beijing, China}
\author{Qizhi~Li}
\thanks{These authors contributed equally to this work}
\affiliation{International Center for Quantum Materials, School of Physics, Peking University, Beijing, China}
\affiliation{Shenzhen Pinghu Laboratory, Shenzhen, China}
\author{Shilong~Zhang}
\affiliation{International Center for Quantum Materials, School of Physics, Peking University, Beijing, China}
\author{Qian~Xiao}
\affiliation{International Center for Quantum Materials, School of Physics, Peking University, Beijing, China}
\author{Sahil~Tippireddy}
\affiliation{Diamond Light Source, Harwell Campus, Didcot, United Kingdom}
\author{Jie~Li}
\affiliation{National Laboratory of Solid State Microstructures and Department of Physics, Nanjing University, Nanjing, China}
\author{Yuchen~Gu}
\affiliation{International Center for Quantum Materials, School of Physics, Peking University, Beijing, China}
\author{Shichi~Dong}
\affiliation{International Center for Quantum Materials, School of Physics, Peking University, Beijing, China}
\author{Taimin~Miao}
\affiliation{Beijing National Laboratory for Condensed Matter Physics, Institute of Physics, Chinese Academy of Sciences, Beijing, China}
\author{Xiangyu~Luo}
\affiliation{Beijing National Laboratory for Condensed Matter Physics, Institute of Physics, Chinese Academy of Sciences, Beijing, China}
\author{Xianghong~Jin}
\affiliation{International Center for Quantum Materials, School of Physics, Peking University, Beijing, China}
\author{Lin~Zhao}
\affiliation{Beijing National Laboratory for Condensed Matter Physics, Institute of Physics, Chinese Academy of Sciences, Beijing, China}
\author{Xingjiang~Zhou}
\affiliation{Beijing National Laboratory for Condensed Matter Physics, Institute of Physics, Chinese Academy of Sciences, Beijing, China}
\author{Ke-Jin~Zhou}
\affiliation{Diamond Light Source, Harwell Campus, Didcot, United Kingdom}
\affiliation{School of Nuclear Science and Technology, University of Science and Technology of China, Hefei, Anhui, China}
\author{Yi~Lu}
\email{yilu@nju.edu.cn}
\affiliation{National Laboratory of Solid State Microstructures and Department of Physics, Nanjing University, Nanjing, China}
\author{Yingying~Peng}
\email{yingying.peng@pku.edu.cn}
\affiliation{International Center for Quantum Materials, School of Physics, Peking University, Beijing, China}
\affiliation{Collaborative Innovation Center of Quantum Matter, Beijing, China}
\author{Yuan~Li}
\email{yuan.li@iphy.ac.cn}
\affiliation{Beijing National Laboratory for Condensed Matter Physics, Institute of Physics, Chinese Academy of Sciences, Beijing, China}
\affiliation{International Center for Quantum Materials, School of Physics, Peking University, Beijing, China}

\date{\today}

\begin{abstract}
\textbf{How electron-phonon interactions influence high-temperature superconductivity in cuprates remains contested \cite{KresinRMP2009,SchriefferBook2007,KeimerNature2015}, and their role outside the CuO$_2$ planes has been largely overlooked. The most conspicuous evidence for such coupling is the ubiquitous 70-meV dispersion kink seen by photoemission \cite{LanzaraNature2001,SobotaRMP2021,YanPNAS2023}, yet its microscopic origin is still debated \cite{DevereauxPRL2004,GiustinoNature2008,HeidPRL2008,IwasawaPRL2008,LiPRL2021}. Here we use oxygen-$K$-edge resonant inelastic X-ray scattering (RIXS) to probe the trilayer cuprate HgBa$_2$Ca$_2$Cu$_3$O$_{8+\delta}$ (Hg1223). When both incident photon energy and polarization are tuned to the apical-oxygen $1s\!\rightarrow\!2p_z$ transition, the RIXS spectra exhibit a ladder of at least ten phonon overtones, evenly spaced by 70 meV, whose intensities follow a Franck-Condon envelope \cite{AmentEPL2011,LeePRL2013,ValePRB2019,FengPRL2020}, signalling exceptionally strong electron-phonon coupling. Quantitative modelling that incorporates core-hole lifetime evaluation yields an apical-phonon coupling energy of 0.25(1) eV, significantly larger than that of the planar stretching mode. Such a coupling strength offers a strong contender for explaining the universal 70-meV kink and suggests that the dominant electron-phonon channel resides outside the CuO$_2$ planes. By elevating inter-layer lattice dynamics from a peripheral factor to a central actor, our results provide a fresh starting point for theories seeking to reconcile strong correlations, lattice dynamics and high-temperature superconductivity.}

\end{abstract}

\maketitle

\pagebreak

It is commonly accepted that the essential Mott physics in cuprates \cite{LeeRMP2006,KeimerNature2015} is captured by quasi-two-dimensional models that retain only the CuO$_2$ planes.  To account for empirical correlation between $T_\mathrm{c}$ and out-of-plane electronic degrees of freedom, most prominently the apical-oxygen 2$p_z$ orbital \cite{DiCastroPRL1991,OhtaPRB1991,PavariniPRL2001,SakakibaraPRB2012}, it is common practice to absorb the latter into effective planar parameters. A prime example is the next-nearest-neighbor hopping $t^\prime$ \cite{PavariniPRL2001}, which strongly influences superconductivity in Hubbard-model calculations \cite{JiangScience2019}. Consequently, searches for an electronically driven pairing interaction have focused on plane-derived electronic structure and interactions \cite{PengNPhys2017,WangNatComm2022,CuiScience2022,WangScience2023}.

Whether the same planar sufficiency applies to electron-phonon interactions remains unsettled. One might argue that because the relevant electronic states are quasi-two-dimensional, the pertinent phonons are likewise confined to the planes. Indeed, intense interest in electron-phonon coupling was triggered by pronounced anomalies observed in electronic spectra \cite{LanzaraNature2001,SobotaRMP2021,YanPNAS2023,LeeNature2006} and lattice dynamics \cite{ReznikNature2006,PengPRL2020,LiPNAS2020,LeeNPhys2021}, commonly attributed to coupling with the in-plane oxygen stretching (half-breathing) mode or the out-of-plane buckling mode of the same planar oxygens.

Decisively isolating phonons that involve apical rather than planar oxygen is experimentally challenging: inelastic neutron or non-resonant X-ray scattering probes all oxygen atoms with identical cross-sections, forcing reliance on comparison to lattice-dynamical calculations whose accuracy is limited by electron correlations and/or charge-order-induced lattice instabilities.  Resonant inelastic X-ray scattering (RIXS) overcomes this limitation by selecting the element and local orbital through resonant conditions defined by the incident photon energy and polarization. However, the method is useful only when the relevant resonant absorption is sufficiently strong, a condition thought unmet for apical oxygens after early work on La$_{2-x}$Sr$_x$CuO$_4$ \cite{ChenPRL1992} suggested their 2$p_z$ states are essentially filled.

Here we exploit O-$K$-edge RIXS to revisit the apical-oxygen phonon problem in the trilayer cuprate HgBa$_2$Ca$_2$Cu$_3$O$_{8+\delta}$ (Hg1223), which holds the highest ambient-pressure $T_\mathrm{c}$ record to date \cite{SchriefferBook2007,KeimerNature2015}.  In Hg1223, well-separated, polarization-dependent absorption peaks allow site-selective excitation of planar or apical oxygen.  By utilizing these conditions for RIXS measurements, we uncover an electron-phonon coupling strength for the apical modes that far exceeds the planar value, rivaling some of the strongest coupling ever reported in the literature. Our findings (i) establish the apical site as the dominant electron-phonon coupling pathway, (ii) highlight the apical-oxygen covalency as a key ingredient for theoretical modelling, and (iii) reopen the discussion of a phononic contribution to the Hg-family of cuprates' exceptionally high $T_\mathrm{c}$.

\newcommand{\subfigref}[2]{\hyperref[#1]{\ref*{#1}#2}}

\subsection*{Unoccupied apical oxygen 2$p_z$ states}

We first present our polarization-dependent O-$K$-edge X-ray absorption spectroscopy (XAS) measurements of Hg1223, which enable us to isolate the resonant conditions of apical and planar oxygen sites. Figure~\ref{figure1}a illustrates our measurement geometry. With the incident electric field ($\mathbf{E}$) set vertical ($\sigma$ polarization), rotating the crystal about the vertical axis leaves the projections of $\mathbf{E}$ onto the crystallographic axes unchanged. The resulting XAS spectra are therefore identical at all incident angles (Fig.~\ref{figure1}b). They display an intense pre-edge peak at $E_i = 529.14$~eV followed by a weaker structure $\sim 2$~eV higher.  For horizontal ($\pi$) polarization, the same rotation sweeps $\mathbf{E}$ from the $ab$-plane toward the $c$-axis direction, producing a clear evolution of the spectra (Fig.~\ref{figure1}c) from normal incidence ($\theta = 90^{\circ}$, identical to the $\sigma$-polarized spectrum in Fig.~\ref{figure1}b) to grazing incidence ($\theta = 20^{\circ}$). The systematic growth of a higher-energy second resonance at $E_i=530.95$~eV with increasing $\mathbf{E}\!\parallel\!c$ immediately associates this feature with orbitals oriented along the $c$ axis. A related peak has been reported in polycrystalline samples \cite{PellegrinPRB1996} but could not be assigned as such.

Figure~\ref{figure1}d makes the linear dichroism explicit by comparing the $\mathbf{E}\!\perp\!c$ and $\mathbf{E}\!\parallel\!c$ components extracted from Figs.~\ref{figure1}b–c.  In the $\mathbf{E}\!\perp\!c$ spectrum, the low-energy peak at about 529 eV peak corresponds to excitation into the Zhang–Rice singlet (ZRS) band, while the hump near 531~eV marks the upper Hubbard band (UHB) \cite{ChenPRL1991,PellegrinPRB1996}. The associated electronic states are formed by the $2p_{x,y}$ and $3d_{x^2-y^2}$ orbitals of the planar O(2)/O(3) and Cu(1)/Cu(2) sites (Fig.~\ref{figure1}e). The additional, purely $\mathbf{E}\!\parallel\!c$ resonance at 530.95~eV is almost absent in La$_{2-x}$Sr$_x$CuO$_4$ \cite{ChenPRL1992} but prominent here. We attribute it to the apical-oxygen O(1) $1s \!\rightarrow\! 2p_z$ transition.  Its strength demonstrates that the $2p_z$ orbital retains substantial unoccupied weight in Hg1223.  In Figs.~\ref{SF1} and \ref{SF2}, we further show that these states exist across a wide range of doping.

Guided by these observations (summarized in Fig.~\ref{figure1}f), we selected the incident energy $E_i=529.14$~eV with $\sigma$ polarization to resonate with planar oxygens and $E_i=530.95$~eV with $\pi$ polarization to target the apical site in the subsequent RIXS measurements. The clear separation of the two resonances enables a direct, site-resolved comparison of phonon excitations, which we discuss next.

\subsection*{Planar and apical oxygen phonons}

Before we present the RIXS data, it is useful to note that when a pronounced phonon signal is seen by RIXS, it already means that the vibrational mode has significant coupling to the electronic states excited by the X-ray photons \cite{AmentEPL2011,GilmorePCCP2023}. This is because the intermediate state is short-lived and locally charge-neutral, prevent it from exerting an indiscriminate impact on the lattice. In other words, the phonon intensity weighs the electron-phonon coupling strength \cite{BraicovichPRR2020}. A more interesting regime is entered when the coupling is very strong: significant atomic displacements take place during the intermediate state's short lifetime, which then get projected into multiple excitations of specific vibrational modes in the electronic deexcited final state. This produces a harmonic progression of phonon peaks in the RIXS spectrum known as overtones.

Let us first examine phonons arising from the planar oxygen (using $E_i = 529.14$~eV, $\sigma$ polarization). By varying in-plane momentum transfer ($q_\parallel$) with sample rotation, we obtain the RIXS intensity map in Fig.~\ref{figure2}a, which shows a dispersing phonon branch that reaches up to $\sim70$~meV. The mode softens to below 50~meV toward $q_\parallel\approx 0.28$ reciprocal lattice units (r.l.u.), where the intensity reaches a maximum.  Since the sample rotation does not affect the resonant condition, the intensity increase points toward an increased electron-phonon coupling strength near the charge-order wave vector \cite{PengPRL2020,LiPNAS2020,LeeNPhys2021}, which is known to be around 0.28~r.l.u. in the Hg-family of cuprates \cite{TabisNatComm2014,WangPRB2020}. This understanding is further supported by observation of extra elastic-line intensity around the same $q_\parallel$, indicative of charge correlations. In spite of this, the RIXS spectrum in Fig.~\ref{figure2}b shows no sign of higher-order replicas above the noise. Based on the observed dispersion, we believe that we are mostly probing the planar stretching mode \cite{LiPNAS2020}. The planar nature is further evidenced as we move the incident energy off-resonance (Fig.~\ref{figure2}c): the RIXS intensity profile is similar to that of the ZRS peak seen by XAS, similar to behaviors of the stretching mode previously observed in bilayer cuprates \cite{PengPRB2022}. In Fig.~\ref{SF3}, we show that the mode can be equally-well observed in a more underdoped sample, also consistent with the previous results.

Switching to the apical-oxygen resonance ($E_i=530.95$~eV, $\pi$ polarization) produces dramatically different results. In the corresponding intensity map (Fig.~\ref{figure2}d), a ladder of phonon overtones is clearly observed above the elastic line, featuring an energy spacing of $\sim70$ meV. The signals exhibit nearly no dispersion with $q_\parallel$, which we have further confirmed for $q_\parallel$ along the diagonal $(h,\!h)$ direction by rotating the crystal about the $c$ axis by $45^\circ$ (Fig.~\ref{figure2}e). Unlike for the planar mode in Fig.~\ref{figure2}a, the decrease of this mode's intensity toward $q_\parallel=0$ is primarily due to the decreasing $\mathbf{E}\!\parallel\!c$ component as $\theta$ increases, which affects the resonant condition (for both incident and scattered photons, see Fig.~\ref{SF4}). Therefore, it is not unlikely that the mode may exhibits similar intensities and strength of the overtones at all $q_\parallel$ if the same resonant condition can be maintained (\textit{e.g.}, by measuring with $\sigma$ polarization on an $ac$ crystal surface). Figure~\ref{figure2}f shows that the RIXS intensity follows the O(1) $2p_z$ XAS peak profile, firmly establishing the mode's distinct origin from the planar mode.

\subsection*{Remarkable electron-phonon coupling strength}

Before interpreting the contrasting results above in terms of electron-phonon interactions, we note that a phonon's dispersion may affect the visibility of its overtones \cite{BraicovichPRR2020}. This is because with a larger dispersion bandwidth, multi-phonon excitations under a total-momentum constraint can occur over a wider range of energy, which reduces the RIXS spectral contrast especially if the range exceeds the energy spacing between the overtones. The opposite limit is a localized, non-dispersing mode, such as in gas molecules \cite{HenniesPRL2010}, which is advantageous for overtone observation. Nonetheless, we argue that the dispersion of the planar mode alone cannot explain its nearly complete lack of overtones. Even with overtone broadening caused by the dispersion, one still expects to see an intensity tail extending toward the high-energy side of the main phonon peak when the overtones are physically present \cite{LeePRL2013,JohnstonNatComm2016,MeyersPRL2018}. In contrast, the RIXS signal terminates very abruptly above the main peak in our data (Fig.~\ref{figure2}b, see also Figs.~\ref{SF3}-\ref{SF5}), followed only by broad photoluminescence signals centered at much higher energies (Fig.~\ref{SF6}).

A simplified model for RIXS phonon overtones \cite{AmentEPL2011,GilmorePCCP2023,RossiPRL2019,ValePRB2019,BraicovichPRR2020} includes an electronic excitation of energy $\epsilon_\mathrm{el}$ (defining the corresponding $E_i$ in XAS), an Einstein phonon of energy $\omega_\mathrm{ph}$, and their coupling energy $M$, as described by the following Hamiltonian:
\begin{equation*}
    \begin{aligned}
    \hat{H} &= \epsilon_\mathrm{el} \hat{c}^\dagger \hat{c} + \omega_\mathrm{ph} \hat{b}^\dagger \hat{b} + M \hat{c}^\dagger \hat{c} (\hat{b}+\hat{b}^\dagger),
    \end{aligned}
\end{equation*}
where $\hat{c}$, $\hat{c}^\dagger$ and $\hat{b}$, $\hat{b}^\dagger$ are the annihilation and creation operators of electronic and phonon excitations, respectively. The electronic excitation is further characterized by a lifetime $1/\Gamma$. Over a wide range of parameters relevant to solids, how overtone intensities decrease with increasing order (known as the Franck-Condon envelope) is controlled by the dimensionless number $M/\Gamma$ \cite{AmentEPL2011,BraicovichPRR2020}, which quantifies the total impact of an excited electron on the lattice during its lifetime. In Fig.~\ref{figure3}a, we present a high-statistics RIXS spectrum for the apical mode to examine the decay of its overtones. The data can be well-fitted to a series of pseudo-Voigt peaks, which account for both the instrumental resolution and the physical energy widths, up to an energy loss of 1 eV with an equal energy spacing set to 70 meV. As shown in the inset, the 10$^\mathrm{th}$-order peak is still clearly visible in the raw data and accurately aligns with the expected energy of 700 meV, indicative of highly harmonic motion at the corresponding vibrational amplitude. Figure~\ref{figure3}b summarizes the fit overtone intensity areas. Using the model above (see Methods), we find a satisfactory description of the Franck-Condon envelope with $M/\Gamma\approx1$. For comparison, $M/\Gamma=0.25$ produces a small $2^\mathrm{nd}$-order signal relative to the $1^\mathrm{st}$ order, close to the detection limit of our $\sigma$-polarization measurements. The two modes thus receive very different impacts from the corresponding electronic excitations on the oxygens.

We emphasize that the above result is expected to be quite robust in spite of the model's simplicity: The apical mode features local vibration, and it is accessed with a well-defined resonant photon absorption which promotes a core electron into a local covalent-bond-like (discussed later) unoccupied state, satisfying all prerequisites of using the simplified model \cite{BraicovichPRR2020}. The extremely clean overtones further rival some of the best examples previously observed in solids and molecules \cite{FengPRL2020,ValePRB2019,LeePRL2013,HenniesPRL2010}. The obtained value of $M/\Gamma\approx1$, however, raises a problem when we combine it with a common practice in the RIXS community, namely, it is often assumed that $\Gamma$ equals 0.15~eV for oxygen core-level excitations \cite{HenniesPRL2010,LeePRL2013,RossiPRL2019,BraicovichPRR2020,PengPRB2022,JohnstonNatComm2016}. Specifically, the chosen value of $\Gamma$ not only translates into $M$ when $M/\Gamma$ is constrained by a successful overtone measurement, as in our case, but also affects the determination of $M$ using an alternative method \cite{BraicovichPRR2020} based on detuning, \textit{i.e.}, how the main phonon peak decreases as $E_i$ is tuned off-resonance. We therefore use our detuning measurement results for the apical mode (Fig.~\ref{figure2}f) to check the common assumption on $\Gamma$. To our surprise, as shown in the inset of Fig.~\ref{figure3}b, simulating detuning with the same model above \cite{RossiPRL2019,BraicovichPRR2020} using $M=\Gamma=0.15$~eV and $\omega_\mathrm{ph}=70$ meV fails to reproduce our observation, which instead suggests $M=\Gamma=0.25(1)$~eV. This value immediately makes $M$ one of the largest in transition-metal oxides \cite{ValePRB2019} (not surprising given the remarkable overtones).

In fact, if we use the width of the XAS peak to estimate $\Gamma$ (Fig.~\ref{SF7}), as has been done in some previous studies \cite{ValePRB2019,RossiPRL2019}, the inferred $M$ is even larger. However, since XAS peaks tend to overestimate $\Gamma$ \cite{WangNatComm2023}, we consider our method above constrained by both overtone and detuning measurements more reliable. Using this method under the constraint of $M=\Gamma/4$, we estimate $M\approx0.122$~eV for the planar mode, \textit{i.e.}, about half the value of the apical mode. The corresponding $\Gamma\approx0.49$~eV is consistent with the width of the XAS peak (Fig.~\ref{SF7}) and considerably larger than that of the apical site, which may reflect how electron itinerancy on the planar orbitals \cite{BraicovichPRR2020,PengPRB2022} is absorbed into our simplified model. Given the dispersion of the planar mode, one might expect the $M=\Gamma/4$ constraint to be further relaxable, and obtain a somewhat larger $M$ and smaller $\Gamma$. Treating these effects (and electronic interactions) at a more rigorous level \cite{ThomasPRX2025} for the planar site will require quantitative information currently unavailable to us, but will unlikely change our conclusion that the apical mode has a significantly larger coupling strength than the planar mode.

While the main body of our results are obtained in nearly optimally doped Hg1223, we have observed very similar behaviors of the apical mode in Hg1223 at a much lower doping (Fig.~\ref{SF8}), as well as in the single-layer compound HgBa$_2$CuO$_{4+\delta}$ (Fig.~\ref{SF9}). The physics therefore appears to be quite universal, at least in the Hg-family of cuprates.

The most direct implication of our results is that the phonon mode with dominant electron-phonon coupling in cuprates might have been misidentified in the past. In particular, the famous 70-meV dispersion kink seen by photoemission \cite{LanzaraNature2001,YanPNAS2023} has long been attributed to the planar stretching mode \cite{DevereauxPRL2004}, but the coupling strength of this mode alone appears insufficient to account for the observed quasiparticle self-energy \cite{GiustinoNature2008,HeidPRL2008}, unless supplemented further by correlation effects \cite{LiPRL2021}. In the light of our discovery, the apical mode naturally offers a simpler, alternative explanation for the kink, which warrants a few remarks here: (i) The apical mode is expected to influence the nodal quasiparticle self-energy by modulating $t^\prime$ \cite{PavariniPRL2001} with its $c$-axis vibration. (ii) In addition to the strong coupling, the mode's flat dispersion near 70~meV also enhances its power for distinctly dressing the quasiparticles. (iii) Unlike the stretching mode, whose energy depends on temperature due to charge fluctuations \cite{LeeNPhys2021}, the apical mode's robust energy \cite{WangPRB2020} is more consistent with the behavior of the kink \cite{YanPNAS2023}. (iv) The apical mode is compatible with oxygen-isotope effect found on the kink \cite{IwasawaPRL2008}. (v) While quasiparticle dispersion anomalies observed in electron-doped cuprates, which have no apical oxygen, appear to support a planar origin \cite{SobotaRMP2021}, the anomalies (especially in the nodal direction) are considerably less pronounced \cite{ShenPMPB2002,ParkPRL2008,HePNAS2019} than the kinks in hope-doped cuprates. It is therefore plausible that the stretching mode only weakly affects the quasiparticle dispersion, while the apical mode is mainly responsible for the kink commonly observed at 70~meV in hole-doped cuprates \cite{SobotaRMP2021} (including in HgBa$_2$CuO$_{4+\delta}$, see Fig.~\ref{SF10}). (vi) In the special case of hole-doped Ca$_2$CuO$_2$Cl$_2$ \cite{RonningPRB2003}, which features heavier apical chlorine atoms, the kink is observed at a significantly lower energy of about 50 meV, even though the CuO$_2$ sheets are the same as in other cuprates.

\subsection*{Covalency and electron correlation}

To further elucidate the nature of the apical mode, we have performed Raman scattering measurements on our Hg1223 crystals. By setting the polarization of incident and scattered photons parallel, we probe $q=0$ phonons under the $A_{1g}$ irreducible representation of the $D_{4h}$ point group. A well-defined peak is observed at 71.8 meV (Fig.~\ref{figure4}a), which can be unambiguously attributed to $c$-axis vibration of apical oxygen. This characteristic mode is also present in the single- and double-layer compounds of the Hg-family of cuprates \cite{WangPRB2020}. Importantly, even though the mode is observed with both $a$- and $c$-axis polarized photons, the Raman intensity is much greater in the $\mathbf{E}\!\parallel\!c$ geometry. This is fully consistent with our findings with RIXS: the Raman process must also involve the $2p_z$ orbitals, in spite of not having a resonantly excited intermediate state. The latter aspect explains why no overtones are observed in the Raman data. To our knowledge, this is the first unambiguous joint observation of an apical phonon mode with both RIXS and Raman spectroscopy.

As an apical oxygen atom moves along the $c$ axis, the most strongly affected (in the percentage sense) inter-atomic distance is from the Hg atom (2~$\mathrm{\AA}$, see Fig.~\ref{figure1}e). In Fig.~\ref{figure4}b, we examine the $2p_z$ unoccupied states on the apical oxygen in conjunction with the Hg orbitals. Using density-functional-theory calculations, we find that the apical-O $2p_z$ and the Hg $5d_{z^2}$ orbitals contribute nearly equally to some of the unoccupied states immediately above the Fermi level, indicative of unsaturated-covalent-like bonding between the two atoms. The covalency implies that the ``spring constant'' between the two atoms, as well as their overall ability to attract electrons from the surrounding, can be strongly influenced by the inter-atomic distance, hence explaining the strong electron-phonon coupling observed in our experiment. Although the RIXS intermediate state involves a core hole, it effectively simulates adding an electron into the apical $2p_z$ orbital and its surrounding, including the planar Cu $3d_{z^2}$ and $4s$ orbitals \cite{PavariniPRL2001}.

The active role of apical oxygen brings a fresh standpoint for theories and exciting new opportunities. The covalency discussed above may explain why material-specific $T_\mathrm{c}$ \cite{PavariniPRL2001,SakakibaraPRB2012} and local pairing strength \cite{OMahonyPNAS2022} correlate with apical-oxygen height, in terms of the lattice's strong static influence on the electronic structure. A dynamic counterpart of this might enable a cooperative interplay between lattice vibrations and spin fluctuations: While spin-mediated interactions have long been seen as the primary glue for Cooper pairs \cite{WangNatComm2022,CuiScience2022,WangScience2023}, certain lattice modes might reinforce the pairing tendencies. Our discovery renders it plausible that the apical oxygen's oscillation dynamically modulates the local electronic environment \cite{PavariniPRL2001,SakakibaraPRB2012,PengNPhys2017} and carrier concentration \cite{KimPRL2018}, in effect working alongside spin fluctuations to stabilize Cooper pairs. Such a dual mechanism---magnetic plus lattice---could help explain novel routes to high-temperature superconductivity \cite{LiuPRX2020} and why the Hg-family of cuprates have exceptionally high $T_\mathrm{c}$. Our findings invite a re-examination of out-of-plane lattice roles in high-$T_\mathrm{c}$ superconductivity, pointing toward a more synergistic picture of the pairing interactions.

\begin{acknowledgments}
We wish to thank T. Devereaux, Kun Jiang, B. Keimer, S.A. Kivelson, B.J. Kim, Dung-Hai Lee, M. Le Tacon, Fa Wang, Nanlin Wang, Hong Yao, and Fuchun Zhang for stimulating discussions, and Yiran Liu for assistance with measurements in related samples. The work is financially supported by the National Key Research and Development Program of China (Grants No. 2021YFA1401900 and No. 2022YFA1403000) and the National Natural Science Foundation of China (Grants No. 12474138, No. 12374143, and No. 12274207).
\end{acknowledgments}

\bibliographystyle{naturemag}
\bibliography{Hg1223}

\pagebreak

\section*{Methods}

{\bf Sample preparation.} Single crystals of Hg1223 were grown with a self-flux method \cite{WangPRM2018}. As-grown samples are underdoped with $T_\mathrm{c}\approx 110$~K, whereas other doping levels are obtained by annealing (Fig.~\ref{SF1}). For the strongly underdoped sample UD69, annealing was conducted at a pressure of 3 $\times$ 10 $^{-2}$ Pa and a temperature of 590 $^\circ$C for 28 hours. The slightly underdoped sample UD130 was annealed in flowing oxygen at 500 $^\circ$C for five days. Prior to loading the samples into the vacuum chambers for RIXS and Raman measurements, the crystals were freshly polished along their $ab$ plane using 0.05 $\mu$m-grade 3M lapping films.

\begin{figure*}[h!]
    \centering
    \includegraphics[width=1\textwidth]{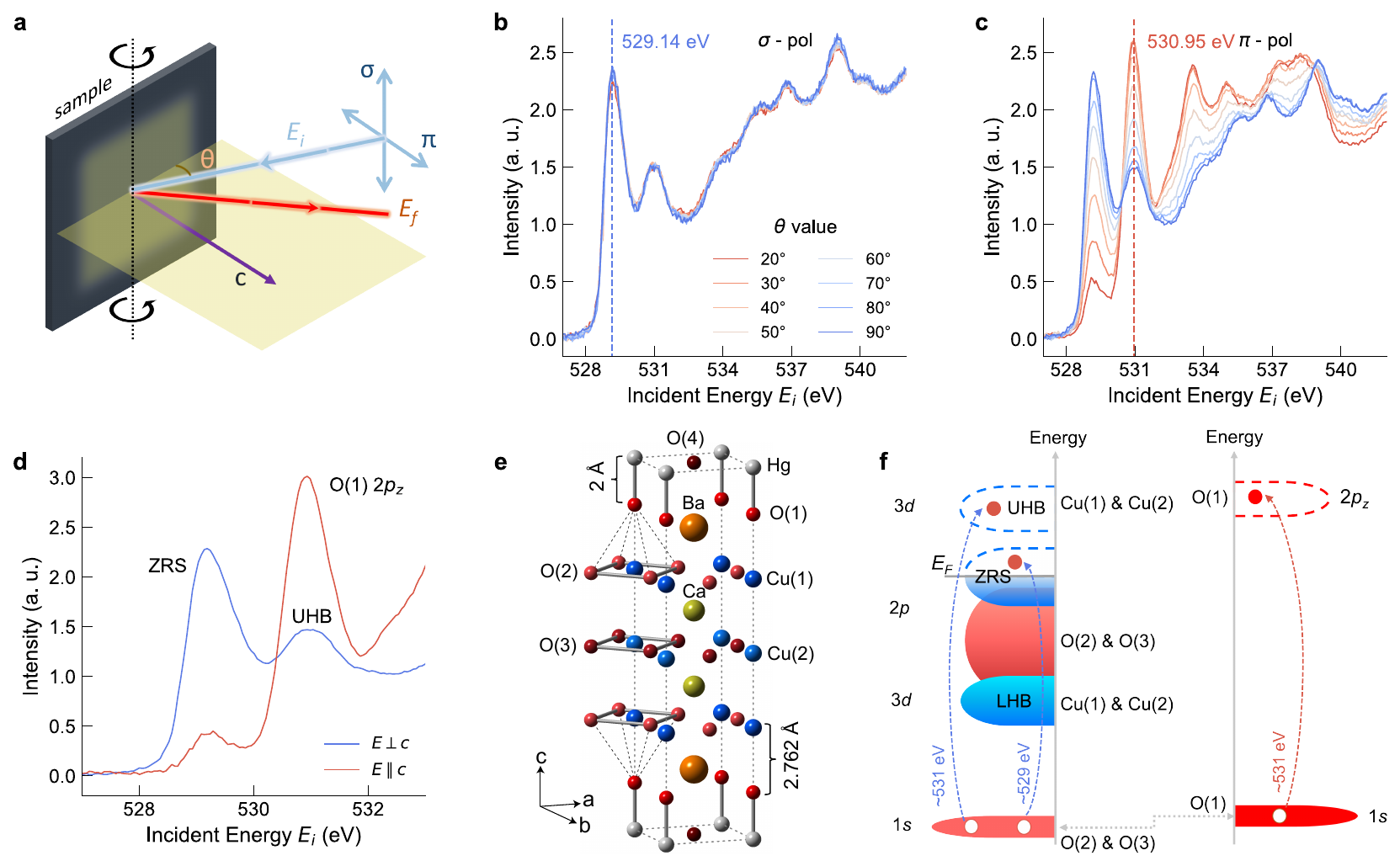}
    \caption{\textbf{XAS signatures of planar and apical oxygen transitions.}
    \textbf{a}, Measurement geometry for XAS (incident beam only) and RIXS (incident and scattered, arranged in a $154^\circ$ back-scattering geometry).  $E_i$ and $E_f$ are the incident- and scattered-photon energies.  The electric field is set either perpendicular ($\sigma$) or parallel ($\pi$) to the scattering plane.  $\theta$ is the incident angle from the crystal surface (perpendicular to the $c$ axis).
    \textbf{b}, $\sigma$-polarized XAS spectra (fluorescence yield) at several $\theta$ values; the curves overlap because $\mathbf{E}\!\perp\!c$ is maintained.
    \textbf{c}, $\pi$-polarised spectra showing the emergence of a peak at $E_i = 530.95$~eV as $\mathbf{E}\!\parallel\!c$ increases. Dash lines in \textbf{b} and \textbf{c} mark the primary conditions chosen for RIXS (Fig.~\ref{figure2}).
    \textbf{d}, Intrinsic $\mathbf{E}\!\perp\!c$ and $\mathbf{E}\!\parallel\!c$ pre-edge spectra obtained by geometric decomposition.
    \textbf{e}, Crystal structure of Hg1223 highlighting planar O(2)/O(3) and apical O(1) sites.
    \textbf{f}, Schematic local density of states.  LHB/UHB: lower/upper Hubbard band; ZRS: Zhang-Rice singlet band.  Vertical dashed arrows indicate in-plane ($\mathbf{E}\!\perp\!c$) and out-of-plane ($\mathbf{E}\!\parallel\!c$) O-$K$-edge transitions associated with planar and apical oxygens, respectively, color-coded with the spectra in \textbf{d}.  Although the O(1) $2p_z$ and UHB features appear at similar absorption energies, their absolute band positions may differ owing to chemical shifts of the core levels (grey dashed arrows).}
    \label{figure1}
\end{figure*}

\begin{figure*}[h!]
    \centering
    \includegraphics[width=1\textwidth]{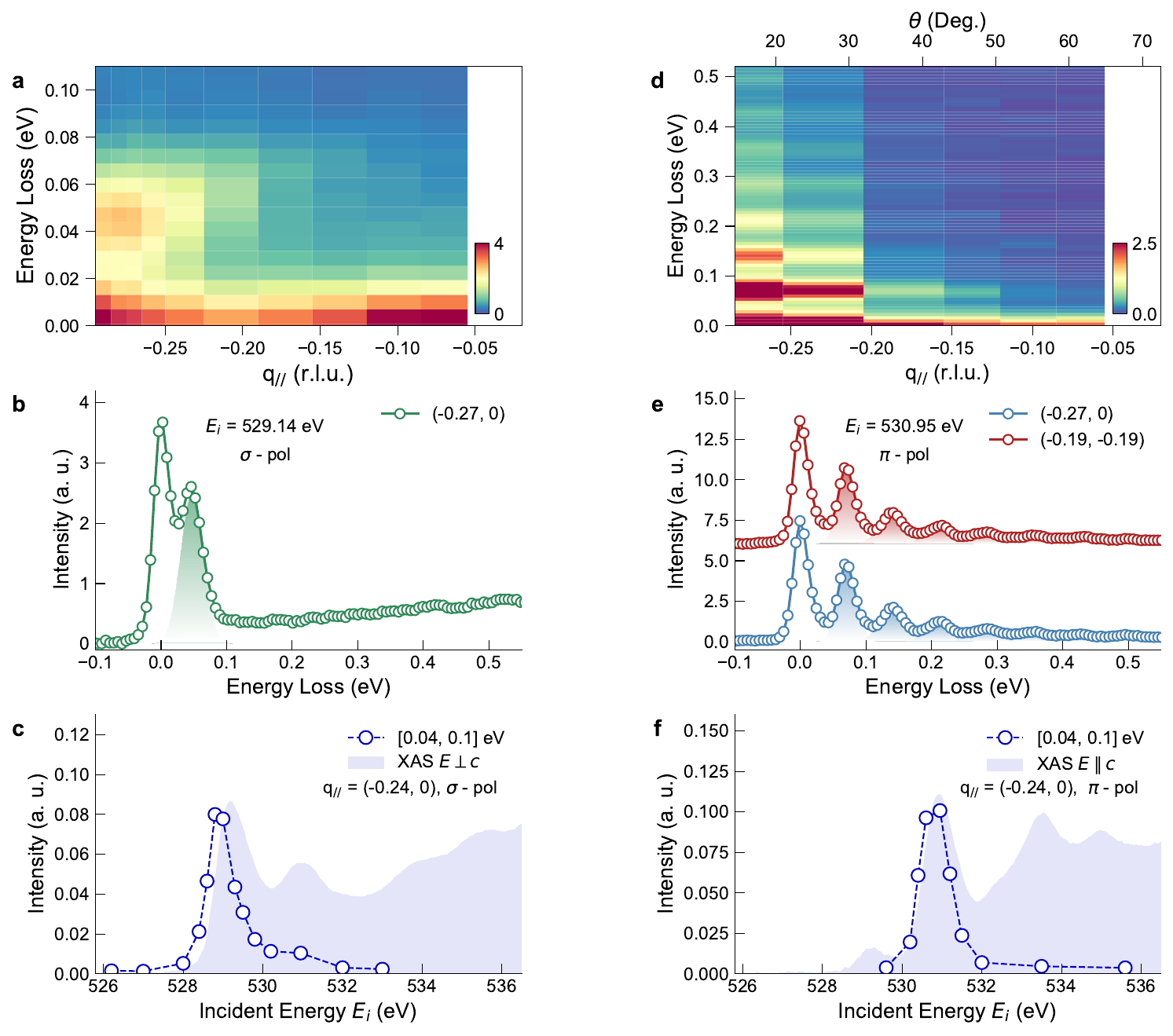}
    \caption{\textbf{RIXS measurements of planar and apical oxygen phonons.}
    \textbf{a}, RIXS intensity map obtained with $E_i = 529.14$~eV and $\sigma$ polarization, probing the planar oxygen site. The in-plane momentum transfer $q_\parallel$ is adjusted by sample rotation, illustrated in Fig.~\ref{figure1}a.
    \textbf{b}, RIXS spectrum taken near the charge-order wave vector (see text).
    \textbf{c}, Incident-energy dependence of the RIXS intensity (integrated from 40 meV to 100 meV, covering the phonon signal), in comparison to the XAS profile for $\mathbf{E}\!\perp\!c$.
    \textbf{d}-\textbf{e}, Same as \textbf{a}-\textbf{b} but for $E_i=530.95$~eV and $\pi$ polarization, probing the apical oxygen site.
    \textbf{f}, Incident-energy dependence of the RIXS intensity, in comparison to the XAS profile for $\mathbf{E}\!\parallel\!c$.
    }
    \label{figure2}
\end{figure*}

{\bf XAS and RIXS measurements.} The XAS and RIXS measurements were performed at beamline I21 of Diamond Light Source, Didcot, United Kingdom. The spectrometer was optimized for the oxygen $K$ edge near 530 eV prior to the measurements, which were performed in the geometry depicted in Fig.~\ref{figure1}b and at a base temperature of 20 K. XAS measurements were performed in both total-fluorescence-yield (TFY) and total-electron-yield modes, and mainly presented in the more bulk-sensitive TFY mode. For RIXS measurements, the instrumental energy resolution was approximately 26 meV (FWHM). The reciprocal lattice units (r.l.u.) for momentum transfer are defined using the lattice constants $a$ = $b$ = 3.86 $\mathrm{\AA}$, $c$ = 15.87 $\mathrm{\AA}$.

\begin{figure*}[h!]
        \centering
        \includegraphics[width=1\textwidth]{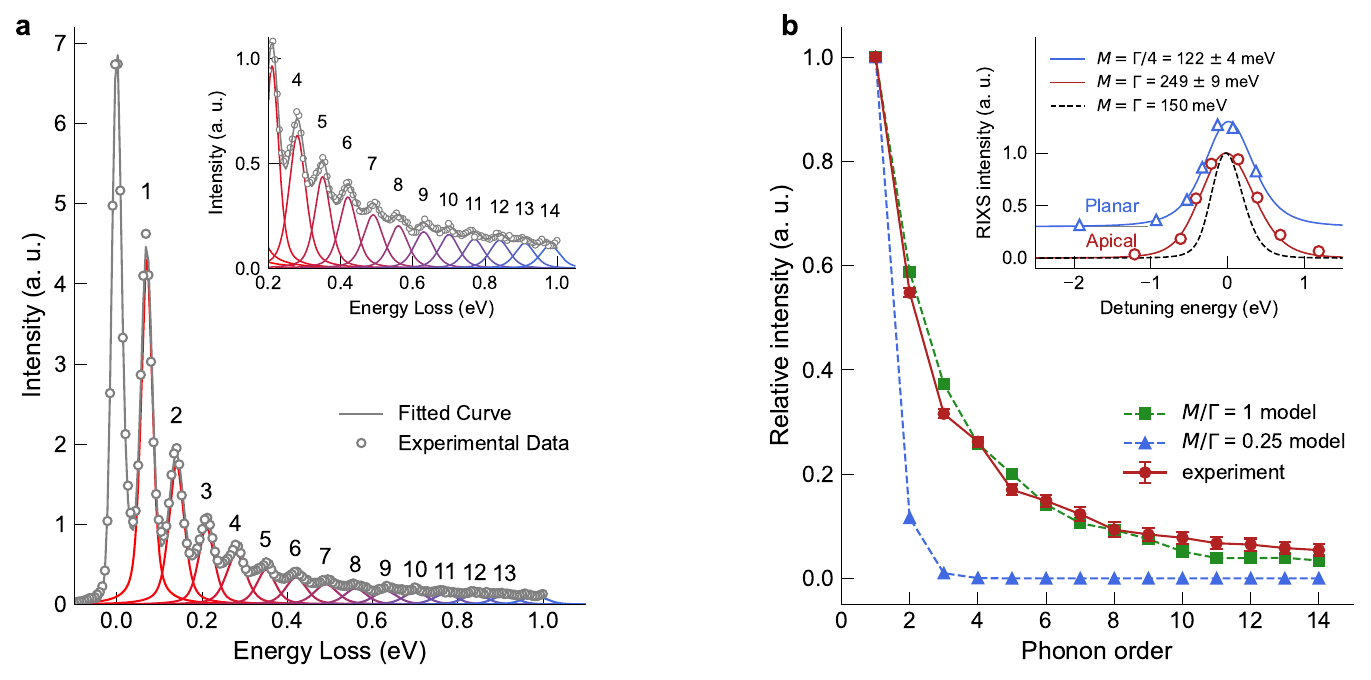}
        \caption{\textbf{Extracting electron-phonon coupling strength from RIXS spectra.}
        \textbf{a}, High-statistics RIXS spectrum measured at $ \theta = 15^\circ $ with $E_i = 530.95$~eV and $\pi$-polarization. Inset shows a zoom-in view of the high harmonics. The data are fit to a series of pseudo-Voigt peaks at equal spacing of 70 meV (see Methods).
        \textbf{b}, Comparison of the fit peak areas in \textbf{a} to model calculations (see text and Methods), normalized to the value of the first order. Inset shows model calculations of the first-order intensity under constraints of $M/\Gamma = 0.25$ and $M/\Gamma = 1$ for the planar and apical modes, respectively, in comparison to the measurement results in Figs.~\ref{figure2}c and \ref{figure2}f.
        }
        \label{figure3}
\end{figure*}

\begin{figure*}[h!]
    \centering
    \includegraphics[width=1\textwidth]{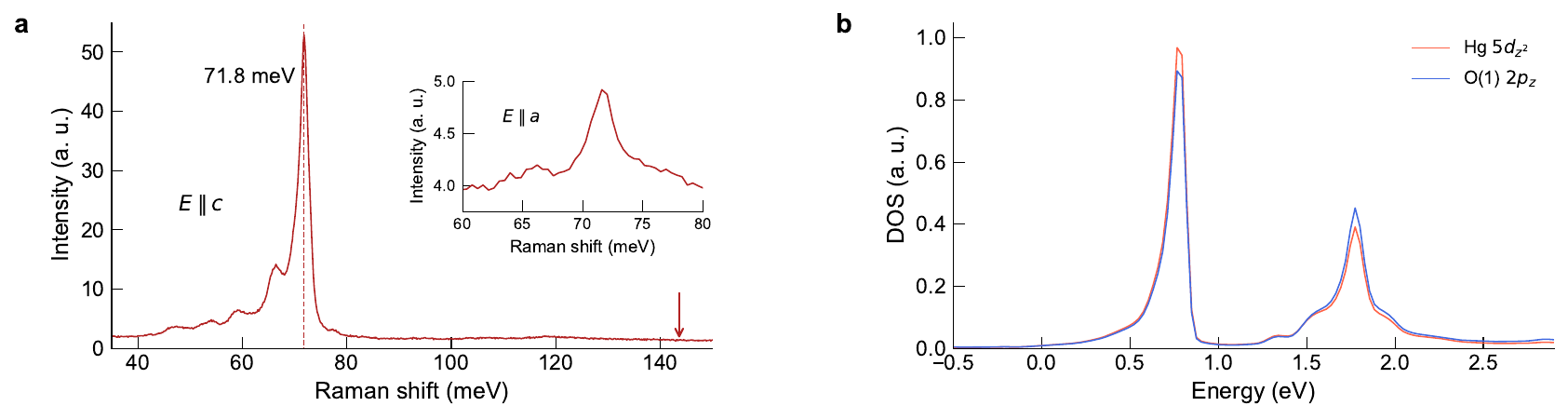}
    \caption{\textbf{Identification of the apical phonon mode and unoccupied $2p_z$ states.}
    \textbf{a}, Raman spectra obtained on a Hg1223 single crystal at 300 K. The measurements are performed with parallel polarization of incident and scattered photons along the $c$ (main) and $a$ (inset) axes. Arrow indicates twice the energy of the main phonon peak at 71.8~meV.
    \textbf{b}, Density-functional-theory calculated partial density of states above the Fermi level arising from the apical O 2$p_z$ and the Hg 5$d_{z^2}$ orbitals.
    }
    \label{figure4}
\end{figure*}

{\bf Data fitting.} The RIXS spectrum in Fig.~\ref{figure3}{a} are fitted to a sum of an elastic component and a progression of phonon overtone peaks defined as:
\begin{equation}
I(x) = \text{pV}_{\text{0}} + \sum_{k=1}^{14} \text{pV}_k,
\end{equation}
where each pseudo-Voigt component $\text{pV}_k(x; A_k, x_k, \gamma, \eta)$ is defined as $A_k[\eta\mathcal{L}(x; x_k, \gamma) + (1-\eta)\mathcal{G}(x; x_k, \gamma)]$, with $\mathcal{L}$ and $\mathcal{G}$ representing Lorentzian and Gaussian profiles. The elastic peak ($\text{pV}_0$) is purely Gaussian ($\eta=0$) and constrained with its center fixed at $x_0 = 0$ eV. The Gaussian full width at half maximum (FWHM) is fitted to be $2\gamma_{\text{el}} = 27.2 \pm 0.2$ meV, very close to the instrumental resolution. Phonon overtone peaks ($k = 1$ to $14$) are fitted with a fixed Lorentzian fraction $\eta = 0.5$ and a fixed energy interval $\omega_\mathrm{ph} = 70$ meV, yielding peak positions $x_k = k\omega_\mathrm{ph}$. The fitted peak FWHMs $2\gamma$ is displayed in Fig.~\ref{SF11}, which increase with increasing order in an approximately linear fashion. Yet, beyond the 9$^\mathrm{th}$ order, it is difficult for a free fit to converge to reasonable results, so we fix the widths at and above the 10$^\mathrm{th}$ order to that of the 9$^\mathrm{th}$ order.

{\bf Calculation of phonon intensities.} Following the notations in \cite{GilmorePCCP2023,BraicovichPRR2020}, the intensity of the $n^\mathrm{th}$ overtone of a phonon mode is calculated as:
\begin{equation}
I_n(\omega_\mathrm{det}) = \big|A_n(\omega_\mathrm{det})\big|^2,
\label{eq:intensity}
\end{equation}
where the scattering amplitude $A_n$ is:
\begin{equation}
A_n(\omega_\mathrm{det}) = \sum_{m=0}^\infty \frac{B_{n,m}(g)B_{m,0}(g)}{\omega_{\mathrm{det}} + (g - m)\omega_\mathrm{ph} + i\Gamma}.
\label{eq:amplitude}
\end{equation}
The Franck-Condon factors $B_{n,m}$ (sorted into the $n\geq m$ order \cite{GilmorePCCP2023}) are:
\begin{align}
B_{n,m}(g) &= (-1)^n \sqrt{\frac{n!m!}{e^{g}}} \sum_{\ell=0}^{m} \frac{(-g)^\ell g^{\frac{n-m}{2}}}{\ell! (m-\ell)! (n-m+\ell)!},
\end{align}
where $g = \left( {M}/{\omega_\mathrm{ph}} \right)^2$ is the dimensionless parameter characterizing the electron-phonon interaction strength, and $\omega_\mathrm{ph}$ is the phonon energy. In the calculation for overtone intensities in Fig.~\ref{figure3}b, we compute the series of $I_n$ with $\omega_\mathrm{det}=0$ for a given set of $M$, $\Gamma$ and $\omega_\mathrm{ph}$. For detuning measurement, the incident energy that maximizes the phonon intensity is used to define $\omega_\mathrm{det}=0$. Relative to this incident energy, the intensity can be computed as \cite{BraicovichPRR2020}:
\begin{equation}
I_1(\omega_{\mathrm{det}}) \propto \left| \sum_{n=0}^\infty \frac{g^n(n-g)}{n!\left[\omega_{\mathrm{det}} + (g - n)\omega_\mathrm{ph} + i\Gamma\right]}\right|^2.
\end{equation}

{\bf Raman scattering experiment.} Raman scattering measurements were conducted on a Horiba Jobin Yvon LabRAM HR Evolution spectrometer in a confocal back-scattering geometry. The setup included a 600 lines/mm grating, a liquid-nitrogen-cooled CCD detector, and a He-Ne laser of $\lambda= 632.8$~nm for excitation. All spectra were acquired at room temperature with incident and scattered photon polarization set to parallel.

{\bf Density-functional-theory (DFT) calculations.} First-principles calculations were performed using the all-electron, full-potential WIEN2K code with the augmented plane-wave plus local orbital (APW+lo) basis set and the Perdew-Burke-Ernzerhof (PBE) exchange functional \cite{BlahaJCP2020,Perdew1996}, on a total number of 1000 $\mathbf{k}$ points.  The density of states (DOS) was calculated by means of the modified tetrahedron method \cite{Andersen1994}. For the DFT+$U$ treatment of Cu 3$d$ electrons \cite{Tran2006}, an effective Hubbard $U$ was chosen as 4 eV in this study.  No significant differences were found between DFT and DFT+$U$ in Hg1223.

\pagebreak
\pagebreak

\pagebreak
\pagebreak

\pagebreak
\pagebreak

\pagebreak
\pagebreak

\pagebreak
\pagebreak

\renewcommand{\thefigure}{S1}
\begin{figure*}[h!]
\centering
\includegraphics[width=0.9\textwidth]{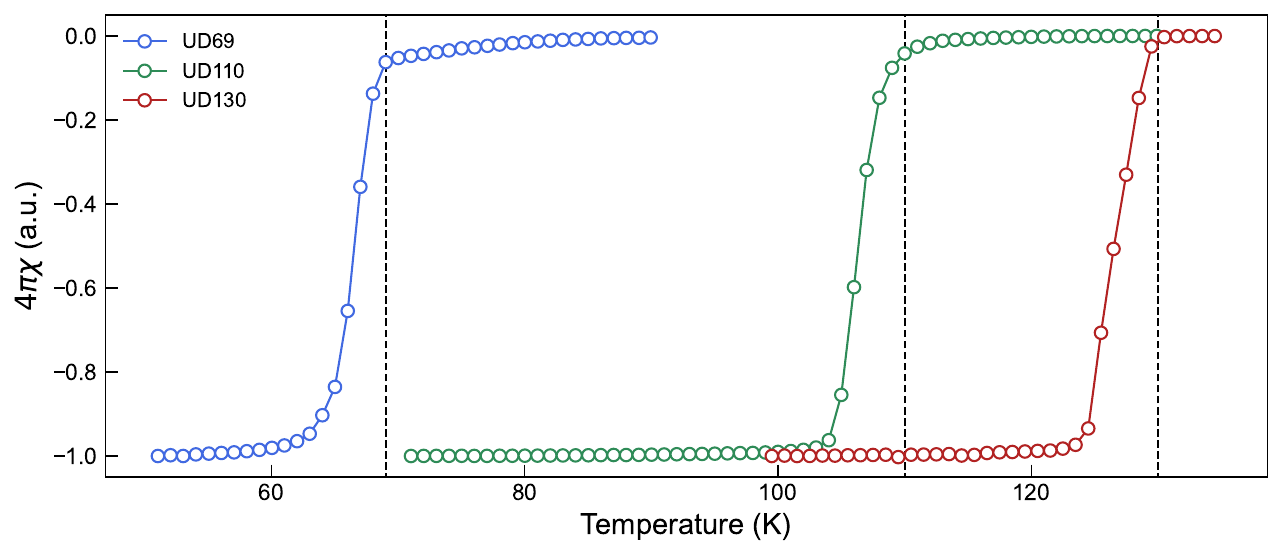}
\caption{\textbf{Hg1223 samples of different doping.}
All samples are underdoped. $T_\mathrm{c}$ values (such as in ``UD130'', in units of kelvin) are defined by the onset of magnetic susceptibility drop. The measurements are done in a magnetic field of 5~Oe along the $c$ axis.
}
\label{SF1}
\end{figure*}

\pagebreak
\pagebreak

\renewcommand{\thefigure}{S2}
\begin{figure*}[h!] \centering
\includegraphics[width=1\textwidth]{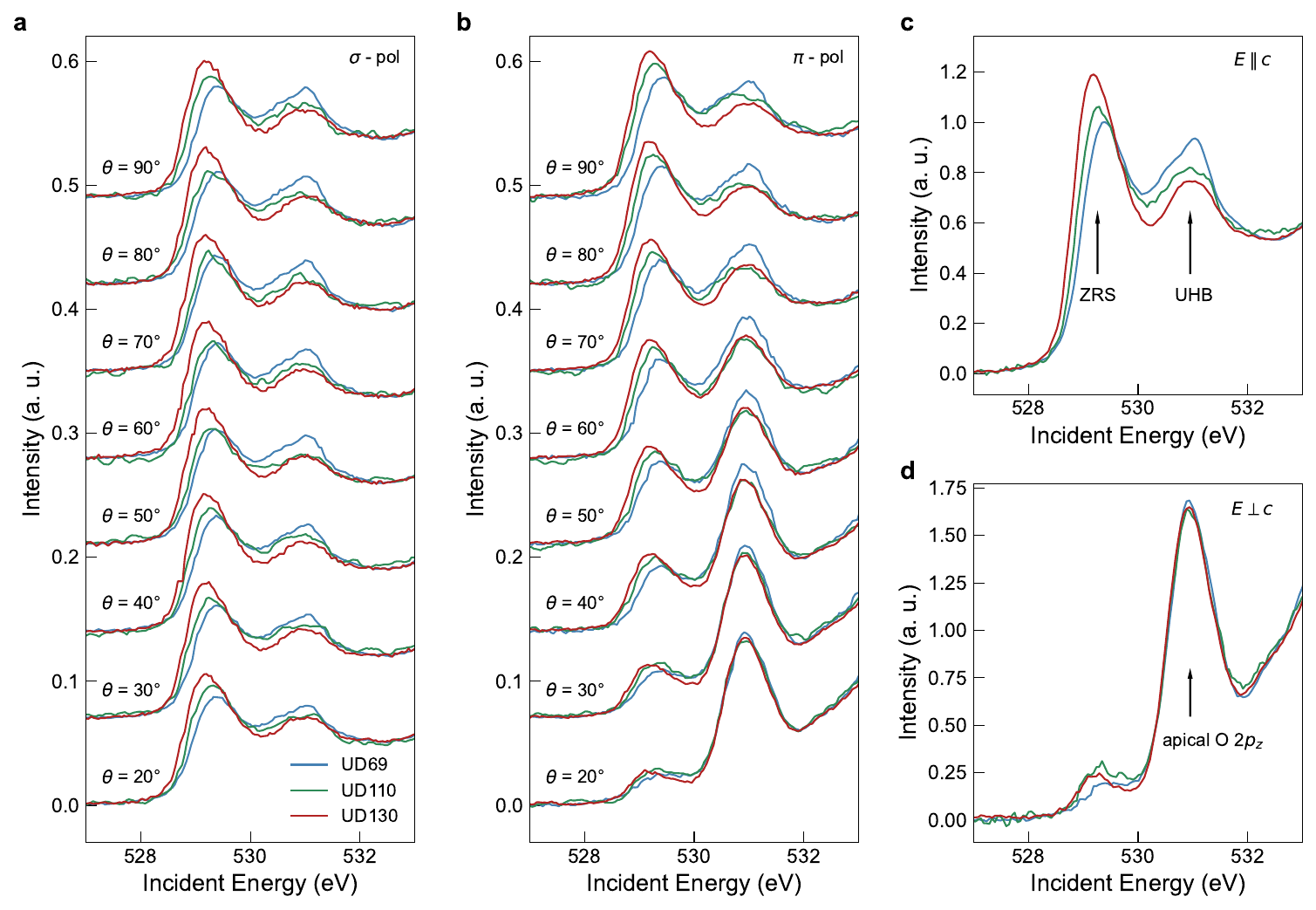}
\caption{\textbf{XAS results obtained at different doping.} \textbf{a} and \textbf{b}, XAS spectra obtained with different incident angles and polarizations, as illustrated in Fig.~\ref{figure1}a. The spectra are normalized at the highest displayed energy and offset for clarity.
\textbf{c} and \textbf{d}, Decomposed in-plane and $c$-axis XAS components.}
\label{SF2}
\end{figure*}

\pagebreak
\pagebreak

\renewcommand{\thefigure}{S3}
\begin{figure*}[h!] \centering
\includegraphics[width=1\textwidth]{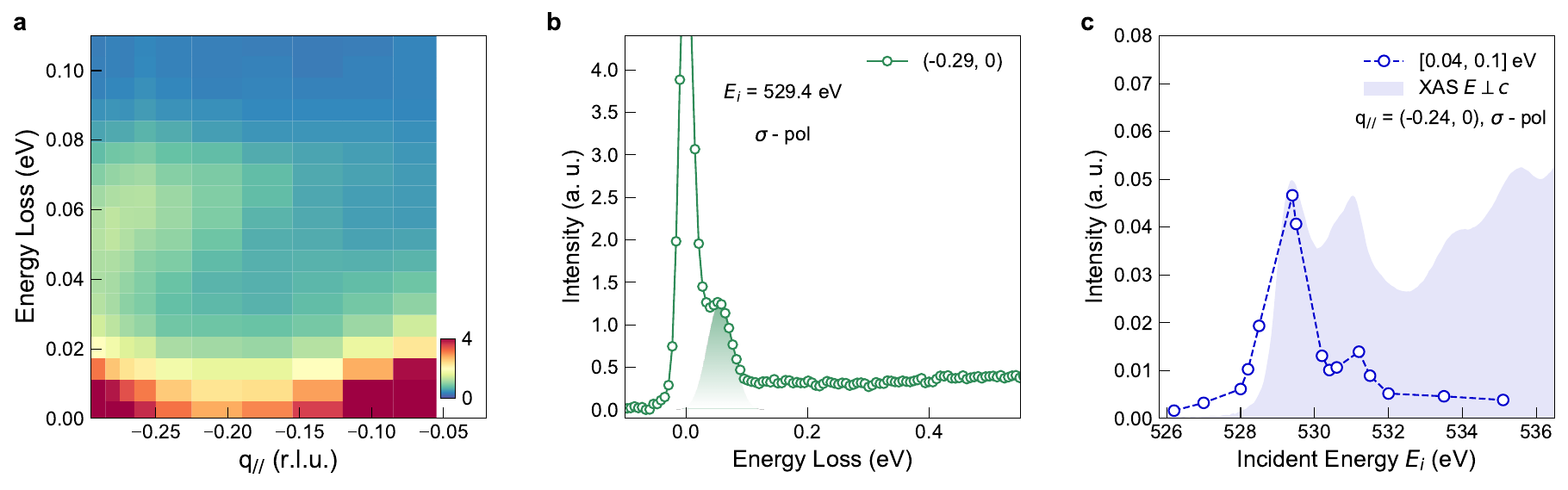}
\caption{\textbf{RIXS results on the planar modes in the UD69 sample of Hg1223.}
Data are presented in a similar form as in Fig.~\ref{figure2}a-c for the UD130 sample.
}
\label{SF3}
\end{figure*}

\pagebreak
\pagebreak

\renewcommand{\thefigure}{S4}
\begin{figure*}[h!] \centering
\includegraphics[width=1\textwidth]{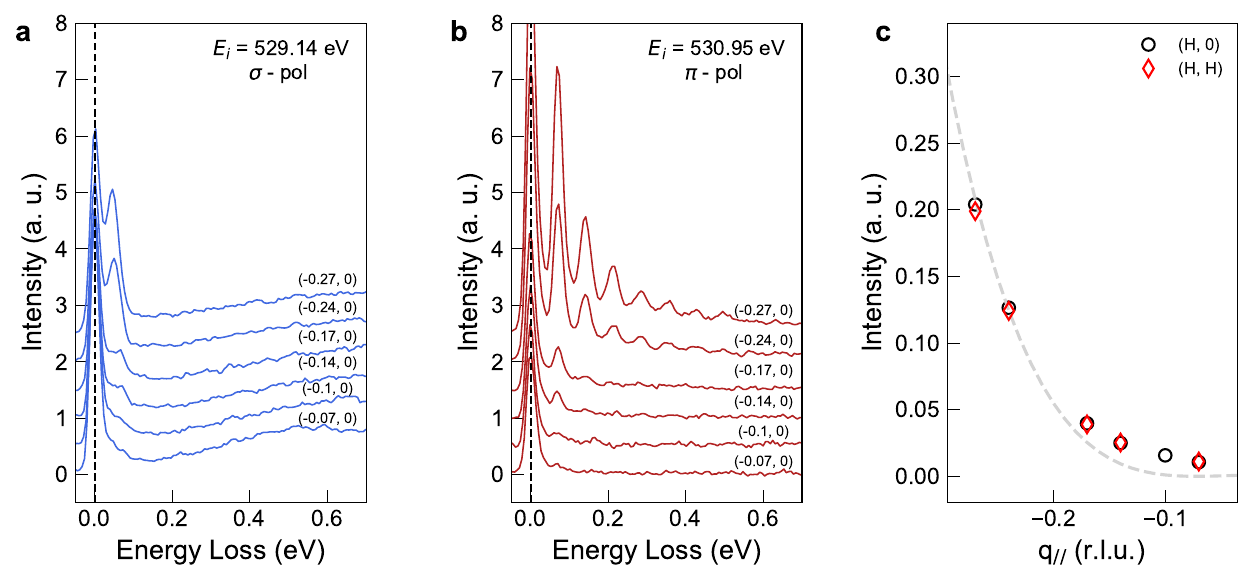}
\caption{\textbf{Detailed RIXS spectra for the UD130 sample of Hg1223.}
\textbf{a} and \textbf{b}, Raw data obtained at different $q_{\parallel}$ on the planar and apical mode, respectively, offset for clarity.
\textbf{c} Integrated apical-mode RIXS intensity in the energy window of [0.04, 0.1]~eV. Dashed line represents the function $y$ = $A \cos^2(\theta) \cos^2(\Omega)$ as a guide to the eye, where $\theta$ and $\Omega$ are incident and scattering angles measured from the sample surface's horizon, respectively, in order to account for the projection $\mathbf{E}\!\parallel\!c$ for both incoming and outgoing photons in the $\pi$ polarization geometry. }
\label{SF4}
\end{figure*}

\pagebreak
\pagebreak

\renewcommand{\thefigure}{S5}
\begin{figure*}[h!] \centering
\includegraphics[width=0.8\textwidth]{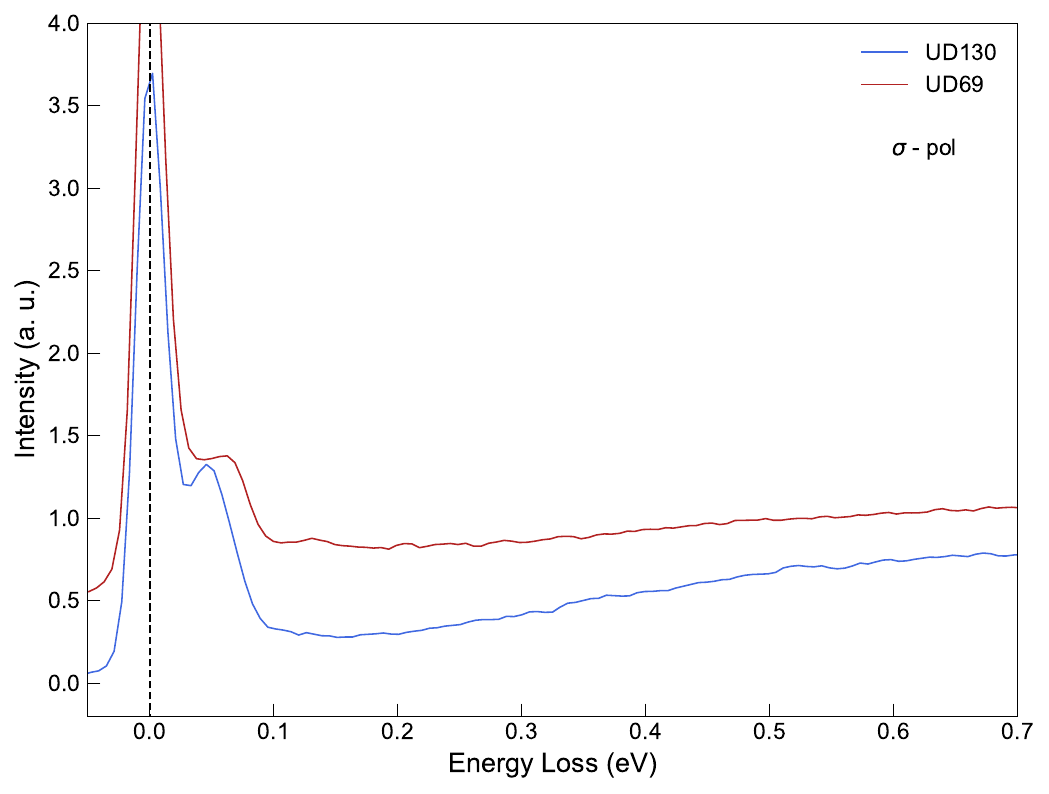}
\caption{\textbf{Momentum-integrated planar-mode RIXS spectra.}
The data are averaged over the spectra displayed in Figs.~\ref{figure2}a and \ref{SF3}a, for the purpose of gaining higher counting statistics to search for overtones.}
\label{SF5}
\end{figure*}

\pagebreak
\pagebreak

\renewcommand{\thefigure}{S6}
\begin{figure*}[h!] \centering
\includegraphics[width=1\textwidth]{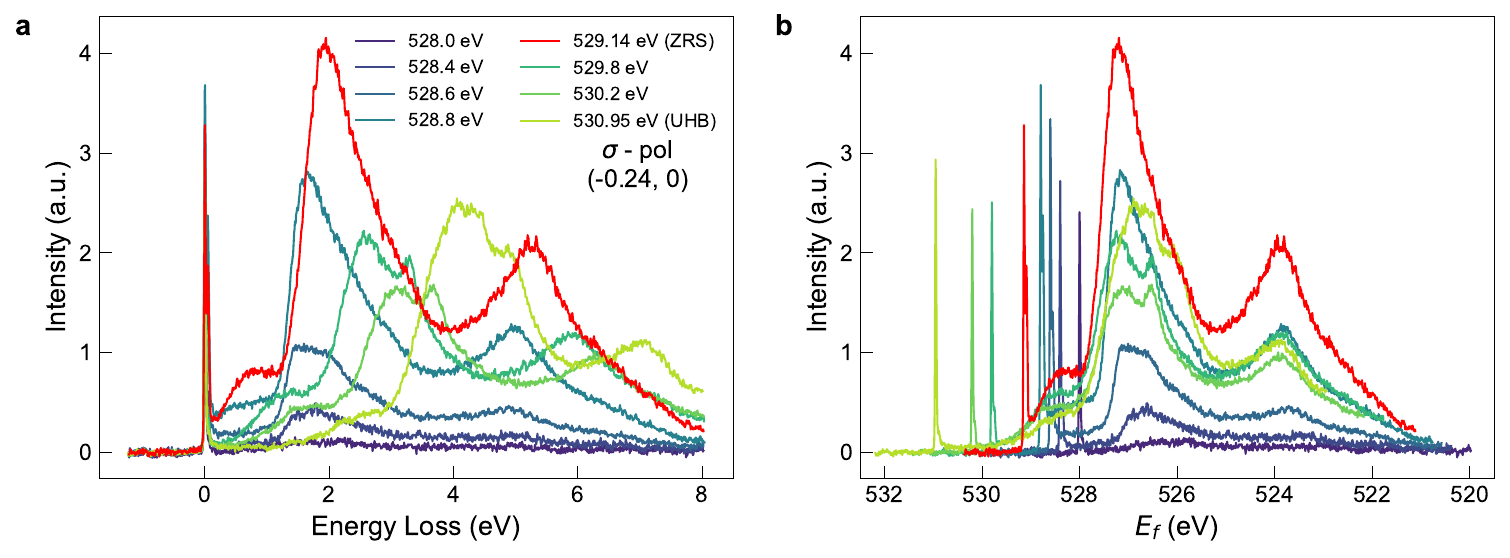}
\caption{\textbf{Photoluminescence in the $\sigma$-polarization RIXS spectra.}
\textbf{a}, Raw data over a wide energy range behind some of the measurements displayed in Fig.~\ref{figure2}c.
\textbf{b}, Same data as in \textbf{a}, but plotted versus the final photon energy. The fact that most of the broad signals are aligned demonstrate that they are of photoluminescence nature.
}
\label{SF6}
\end{figure*}

\pagebreak
\pagebreak

\renewcommand{\thefigure}{S7}
\begin{figure*}[h!]
\centering
\includegraphics[width=1\textwidth]{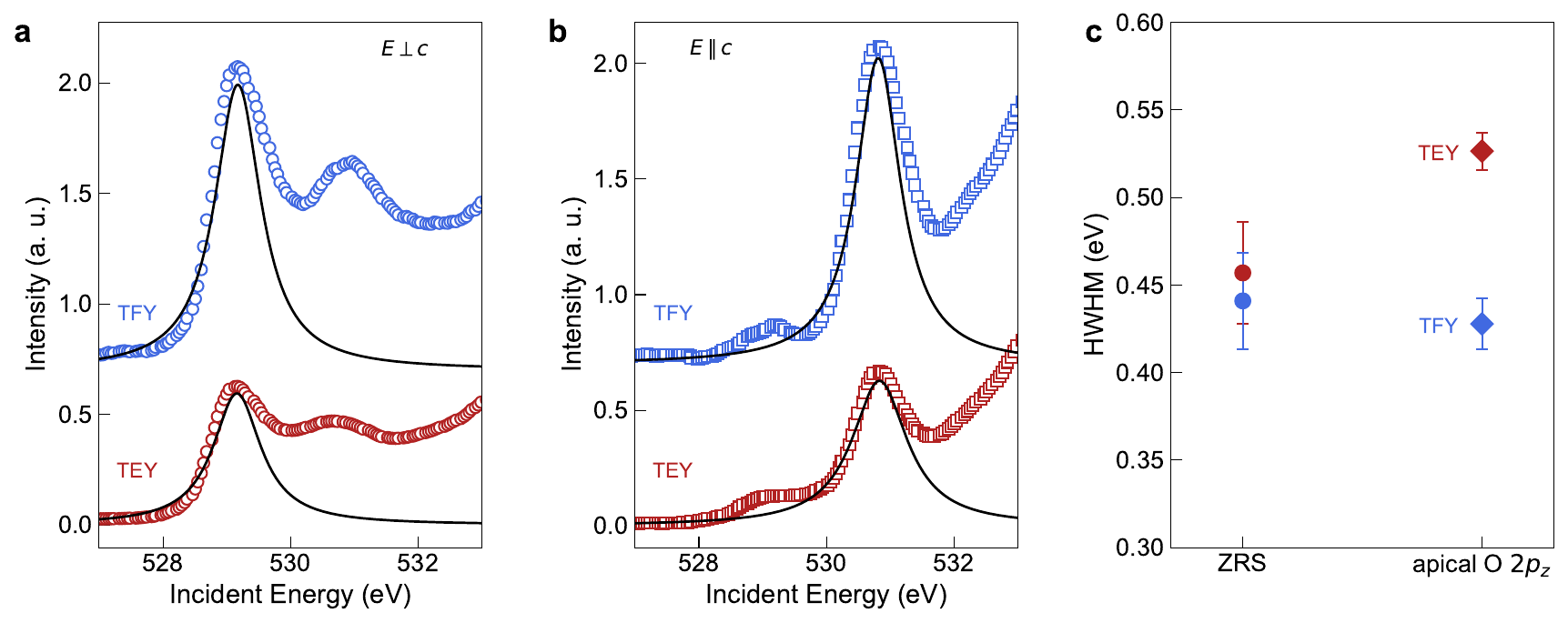}
\caption{\textbf{Energy-width analysis of the resonances used for RIXS.}
\textbf{a} and \textbf{b}, normalized XAS spectra in the pre-edge energy range. The data are fitted with three Lorentzian peaks. Only the ones pertinent to the ZRS and the $2p_z$ resonances are displayed. The Lorentzian line shape is not entirely satisfactory but still suitable for a crude estimate of the energy widths, which are summarized in \textbf{c}.}
\label{SF7}
\end{figure*}

\pagebreak
\pagebreak

\renewcommand{\thefigure}{S8}
\begin{figure*}[h!] \centering
\includegraphics[width=1\textwidth]{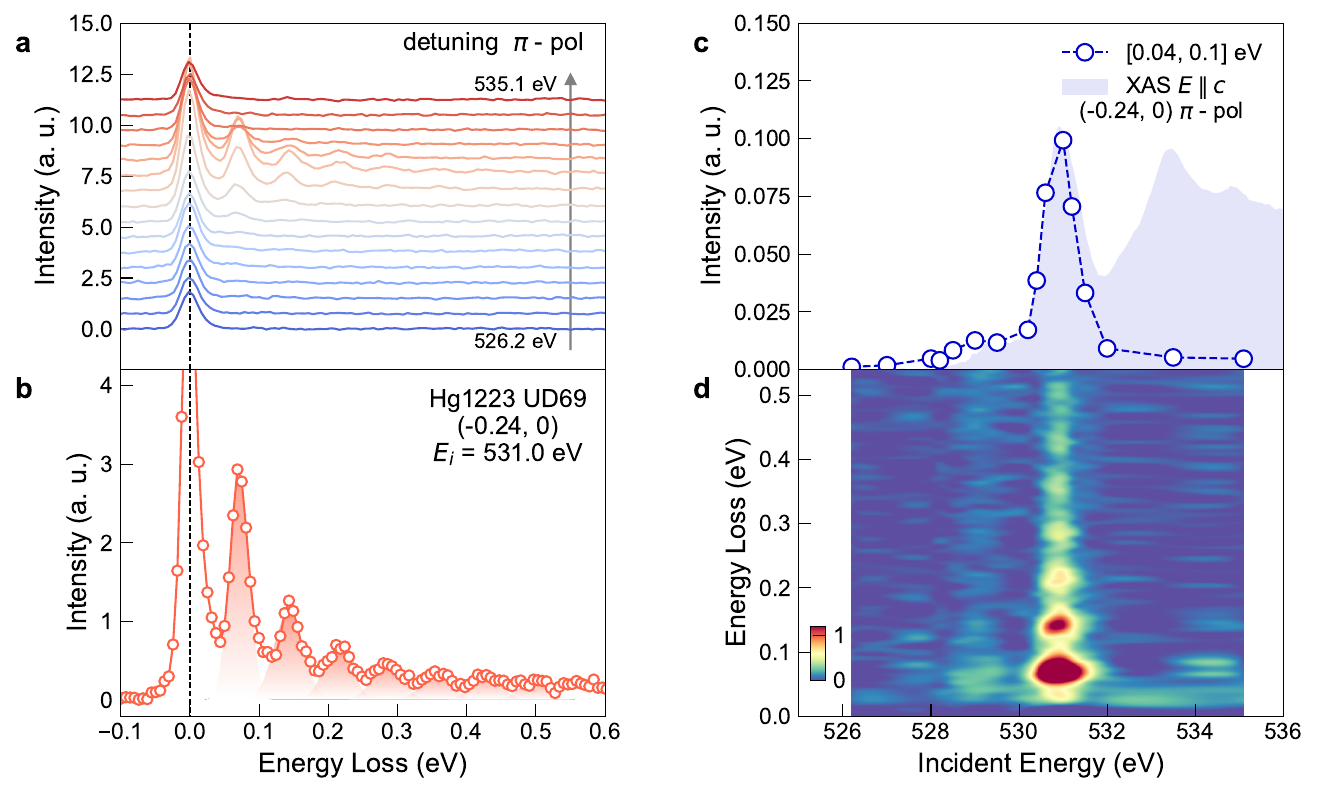}
\caption{\textbf{RIXS spectra of the apical mode in the UD69 sample of Hg1223.} All data are for $q_\parallel=(0.24, 0)$~r.l.u. \textbf{a}, Spectra acquired at different incident energies. \textbf{b}, Spectrum acquired on-resonance at $E_i =531.0$~eV, which shows the most pronounced phonon overtones. \textbf{c}, Detuning behavior of the 1$^\mathrm{st}$-order phonon intensity in comparison to XAS data. \textbf{d}, RIXS intensity map constructed from the data in \textbf{a} after subtracting the elastic line.
}
\label{SF8}
\end{figure*}

\pagebreak
\pagebreak

\renewcommand{\thefigure}{S9}
\begin{figure*}[h!] \centering
\includegraphics[width=1\textwidth]{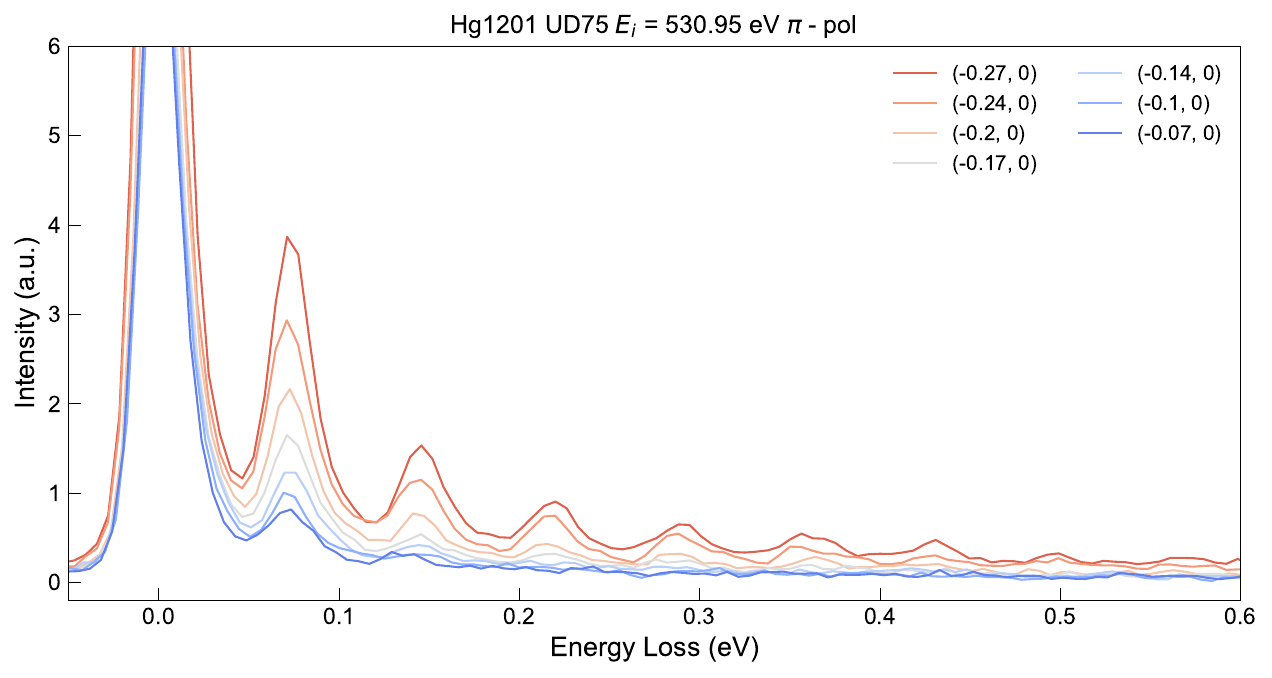}
\caption{\textbf{RIXS spectra of the apical mode in HgBa$_2$CuO$_{4+\delta}$.} The sample is underdoped, with $T_\mathrm{c}=75$~K, prepared with methods described in Refs.~\cite{BarisicPRB2008,ZhaoAdvMater2006}. Measurement conditions and $q_\parallel$ are indicated in the figure.
}
\label{SF9}
\end{figure*}

\renewcommand{\thefigure}{S10}
\begin{figure*}[h!] \centering
\includegraphics[width=1\textwidth]{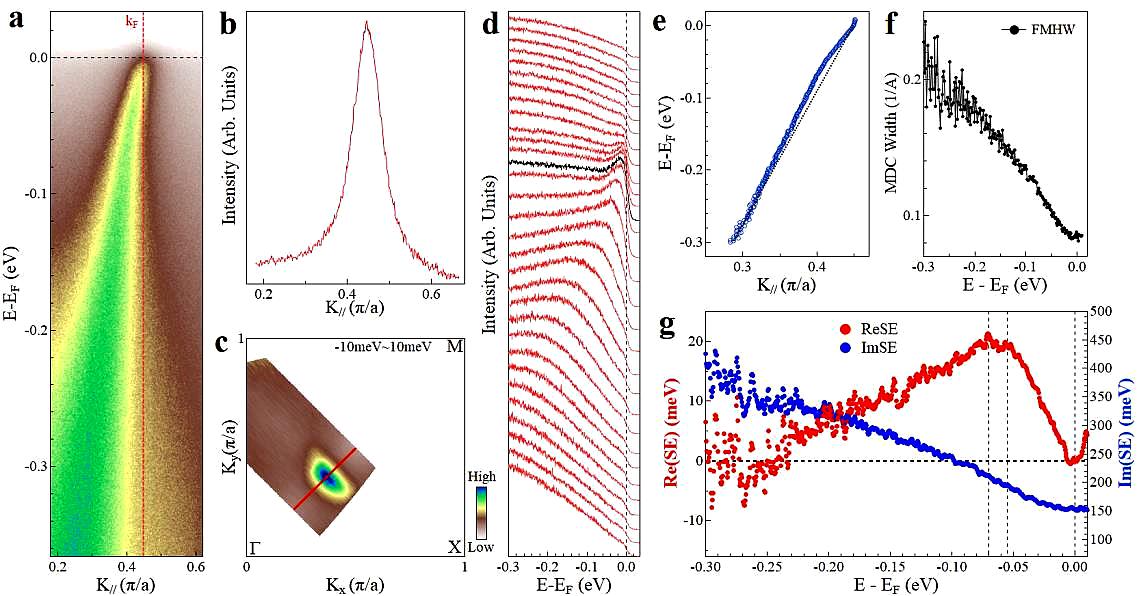}
\caption{\textbf{ARPES data for HgBa$_2$CuO$_{4+\delta}$ in the nodal direction.}
The sample is underdoped, with $T_\mathrm{c}=77$~K.
\textbf{a}, Band structure. Red dashed line marks the Fermi momentum position $k_\mathrm{F}$.
\textbf{b} and \textbf{c}, Momentum distribution curve (MDC) and Fermi surface, integrated over $\pm10$~meV about the Fermi energy.
\textbf{d}, Energy distribution curve (EDC) of each momentum position along nodal direction, offset for clarity.
\textbf{e}, Band dispersion extracted from MDC peak positions.
\textbf{f}, MDC width versus energy.
\textbf{g}, Real and imaginary parts of the self-energy. The kink in \textbf{e} can be identified as two nearby anomalies marked by vertical dashed lines. The one at about 70 meV is ubiquitously observed in hole-doped cuprates \cite{SobotaRMP2021}. The other one close to 50 meV is consistent with a previous report \cite{VishikPRB2014}. The figure is reproduced from Dr. Xiangyu Luo's PhD thesis \cite{LuoPhD2023}, Fig.~5-4, available at DOI:10.27604/d.cnki.gwlys.2023.000129.
}
\label{SF10}
\end{figure*}

\renewcommand{\thefigure}{S11}
\begin{figure*}[h!] \centering
\includegraphics[width=0.8\textwidth]{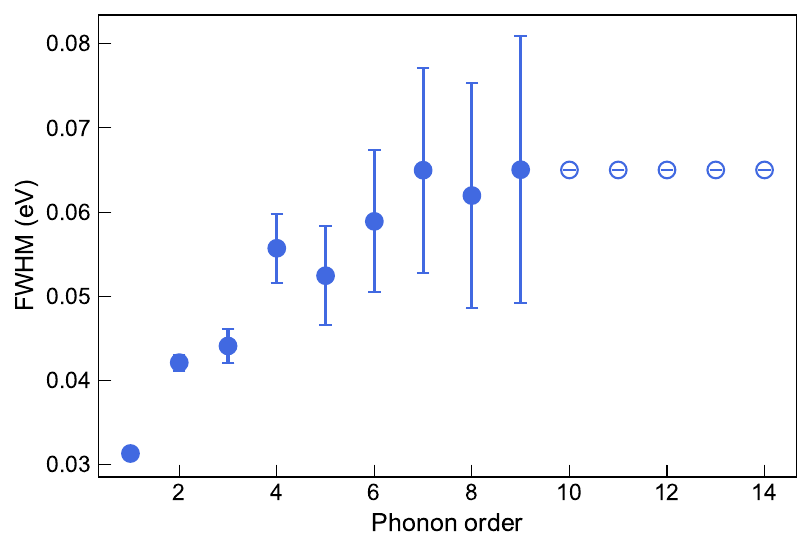}
\caption{\textbf{Fitted peak FWHM of the phonon overtones.} Definition and procedure of the fitting are described in Methods. Values of FWHM at the 10$^\mathrm{th}$ order and above are fixed to be the same as the fitted value of the 9$^\mathrm{th}$ order.
}
\label{SF11}
\end{figure*}

\end{document}